\documentclass[aps,pra,twocolumn,superscriptaddress]{revtex4-1}
\usepackage{bm}
\usepackage{textcomp}
\usepackage{amsmath}
\usepackage{graphicx}
\usepackage{dcolumn}
\usepackage[utf8]{inputenc}

\begin{document}

\title{Rotation of a liquid crystal by the Casimir torque}

\author{David A.T. Somers}
\affiliation{Department of Physics, University of Maryland, College Park, MD 20740, USA}
\affiliation{Institute for Research in Electronics and Applied Physics, University of Maryland, College Park, MD 20740, USA}

\author{Jeremy N. Munday}
\affiliation{Institute for Research in Electronics and Applied Physics, University of Maryland, College Park, MD 20740, USA}
\affiliation{Department of Electrical and Computer Engineering, University of Maryland, College Park, MD 20740, USA}

\date{\today}
\date{\today}

\begin{abstract}
We present a calculation of the Casimir torque acting on a liquid crystal near a birefringent crystal. In this system, a liquid crystal bulk is uniformly aligned at one surface and is twisted at the other surface by a birefringent crystal, e.g. barium titanate. The liquid crystal is separated from the solid crystal by an isotropic, transparent material such as SiO$_2$. By varying the thickness of the deposited layer, we can observe the effect of retardation on the torque (which differentiates it from the close-range van der Waals torque). We find that a barium titanate slab would cause 5CB (4-cyano-4$'$-pentylbiphenyl) liquid crystal to rotate by 10$^\circ$ through its bulk when separated by 35 nm of SiO$_2$. The optical technique for measuring this twist is also outlined.
\end{abstract}


\maketitle

\section{Introduction}
In 1948, Casimir \cite{Casimir1948} calculated that quantum fluctuations of electromagnetic fields cause attraction between two parallel metal plates at zero temperature. This can be interpreted as a physical manifestation of the zero-point energy predicted by quantum field theory \cite{Milonni1994a}. As the plates are brought near each other, the allowed modes between them are limited. Because each electromagnetic mode has an energy of $\hbar \omega/2$ at $T=0$ K, the total energy of the configuration depends on the plate separation. This energy landscape results in an attractive force between the plates.

While Casimir's description holds for ideal metal plates, a more general expression was derived by Lifshitz, Dzyaloshinskii, and Pitaevskii \cite{Lifshitz1956,Dzyaloshinskii1961}. This formulation is used for comparison with experiments and is valid for both real metals and dielectrics. In this theory, the force between two uncharged surfaces can be derived according to an analytical formula (often called the Lifshitz or Casimir-Lifshitz equation) that relates the zero-point energy to the dielectric functions of the interacting surfaces and of the medium in which they are immersed. In 1972, Parsegian and Weiss \cite{Parsegian1972} derived an expression for the nonretarded (where the speed of light is taken to be infinite) interaction energy between two dielectrically anisotropic plates immersed in a third anisotropic material. Barash \cite{BarashRussian} analyzed a similar problem including retardation effects and found an equation for the Helmholtz free energy per unit area. In the nonretarded limit, the results of Parsegian, Weiss, and Barash are in agreement.

In the expressions derived by Parsegian, Weiss, and Barash, the Casimir energy between two parallel, birefringent plates depends on both their separation and their orientation. Between two parallel, positive birefringent slabs with in-plane optical anisotropy, the lowest energy state is when the two optical axes are aligned. This results in a torque that rotates the plates to this configuration.

While the Casimir-Lifshitz force is the subject of much discussion and has been verified in a number of experiments \cite{Lamoreaux1997a,Munday2009,Decca2005,Chan2001,Bressi2002,Mohideen1998}, there have been no published experimental attempts to measure the torque between anisotropic materials. In addition to the early work of Parsegian, Weiss and Barash, recent theoretical work has been performed including a derivation of a more simplified equation of the torque between two plates in one dimension \cite{Enk1995}, the torque between two dielectric slabs with different directions of conductivity \cite{Kenneth2001}, and numerical calculations based on real materials \cite{Munday2005,Munday2006}.

Most discussions of potential Casimir torque measurements involve either levitating microdisks \cite{Munday2005,Munday2006} or a torsion pendulum \cite{Chen2011}. In the first method, microscale birefringent disks would be levitated over a birefringent substrate with Casimir-Lifshitz repulsion or an electrostatic force. The Casimir torque would then cause the freely rotating disks to rotate so that their optical axes would align with that of the substrate. In the proposed torsion pendulum experiment, a macroscopic birefringent crystal would be attached to a quartz torsion pendulum and brought near another birefringent crystal. The Casimir torque would affect the period of natural oscillations of the torsion pendulum, allowing for its detection.

Here we propose a method that is in analogy to a static torsion pendulum with a thick liquid crystal layer as the twisted bulk. As the uniformly aligned liquid crystal is brought near a birefringent crystal, the Casimir torque aligns the liquid crystal molecules with the solid crystal's optic axis, which in turn causes a twist through the bulk of the liquid crystal. A similar experiment was proposed by Smith and Ninham \cite{Smith1973} in 1973 but, to our knowledge, was never carried out. Here we provide a calculation including retardation effects of the expected results of a similar geometry, as well as detailed experimental considerations.

\begin{figure}
	\fbox{\includegraphics[width=0.45\textwidth]{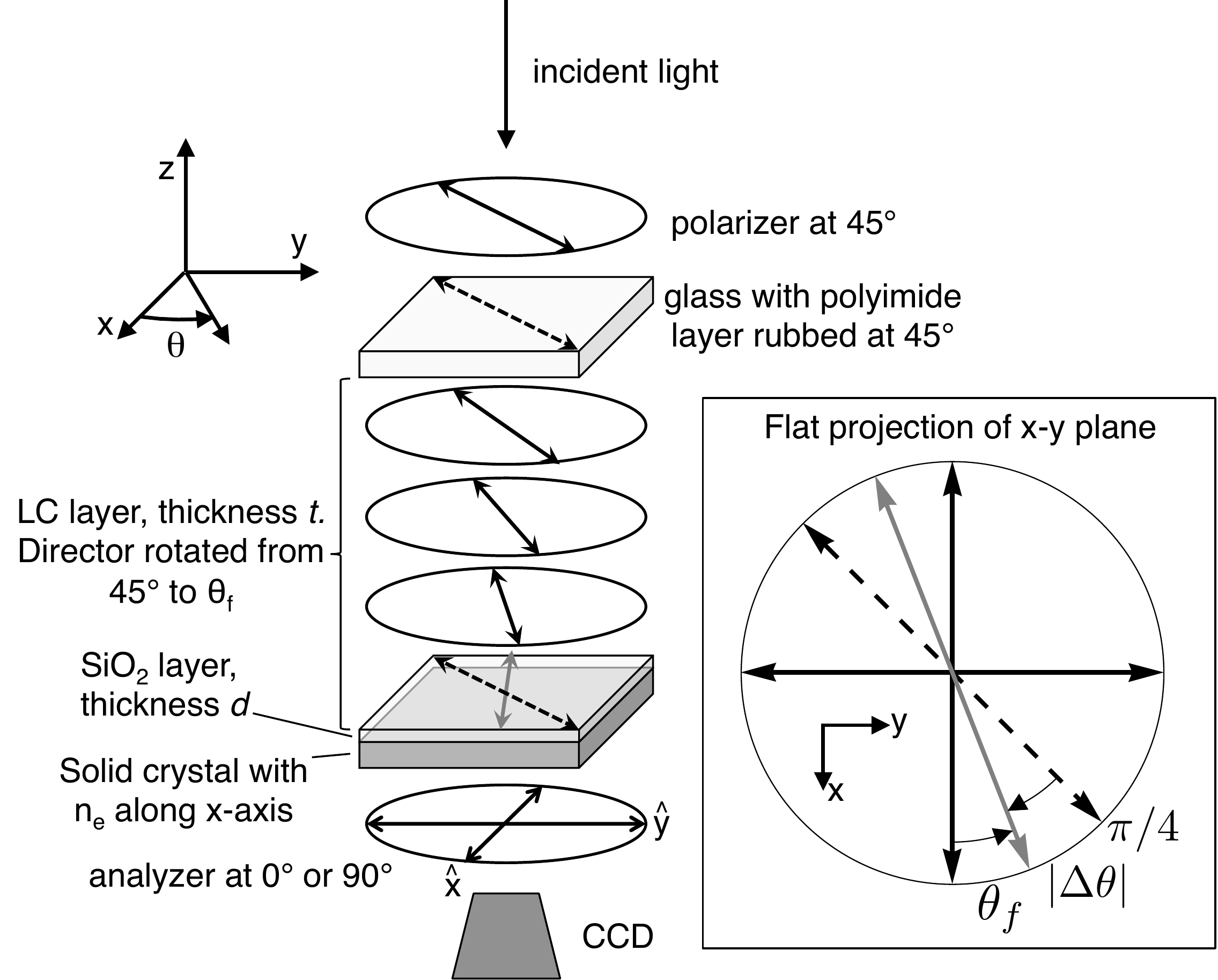}}
	\caption{\label{fig:setup}Experimental setup. The liquid crystal director is fixed at $45^\circ$ at the top surface, but a Casimir torque at the bottom surface causes a linear twist throughout the bulk. Incident light polarized at $45^\circ$ is adiabatically twisted with the liquid crystal director to a final polarization state, $\theta_f$, which can be measured optically. Stronger Casimir torques cause a greater director twist. The inset shows a flat projection of the $x$-$y$ plane.}
\end{figure}

Instead of using the liquid crystal in the isotropic phase as a spacer layer (which is experimentally unfeasible), a thin layer of SiO$_2$ separates the liquid crystal from the solid crystal. Varying the SiO$_2$ thickness is equivalent to changing the distance between parallel plates. If both the liquid crystal and the birefringent crystal have positive uniaxial anisotropy, then the Casimir torque causes the liquid crystal molecules to twist towards the extraordinary axis of the solid crystal. Because the liquid crystal is anchored at the glass interface, the director (i.e. the locally preferred molecular orientation) is twisted through the bulk by the Casimir torque at the boundary at the opposite interface. This geometry is depicted in Fig. \ref{fig:setup}. This experimental design is similar to methods for measuring the azimuthal anchoring energy of liquid crystals on treated substrates \cite{Faetti2003}.
The mechanism of liquid crystal alignment induced by a rubbed polymer layer itself is the subject of much study but is thought to include physical grooves, aligned polymer chains, and the van der Waals torque for surfaces in contact \cite{Nishikawa1998,Ishihara2005a,Wu1996a,Pizzirusso2012}. The last of these is equivalent to the short-ranged Casimir torque of our proposed study, though distinguishing the van der Waals effect from other alignment effects is difficult. However, previous experiments have suggested the anisotropic van der Waals effect as the mechanism of liquid crystal alignment at a surface. Schadt and Schmitt \cite{Schadt1992} used linearly photopolymerized layers to align liquid crystals with a van der Waals interaction. Lu \cite{Lu2004} also provided evidence that the van der Waals interaction is an important component of the liquid crystal alignment at treated polymer layers. However, these experiments do not isolate the van der Waals torque from other surface effects, because the liquid crystal is in contact with the substrate. Our proposed experiment would demonstrate this effect over a distance of tens of nanometers (and in doing so measure the long-range retardation effects of the Casimir torque). Finally, we can relate measured data to Casimir torques calculated from the dispersive properties of the materials.

Smith and Ninham \cite{Smith1973} considered the nonretarded case of this system and predicted measurable distortions of the liquid crystal director. Here we carry out the full retarded calculation of the Casimir torque by considering its effect on a thin boundary layer. That is, in comparison to the liquid crystal bulk with thickness $t>50~\mu$m, most of the Casimir torque is felt by a thin layer of thickness $\delta t<50$ nm. Also, because the total liquid crystal twist through the bulk is less than $45^\circ$, the liquid crystal in the region of $\delta t$ is nearly uniformly aligned. Therefore, we approximate the Casimir torque on this layer from the uniaxial crystal to be the same as that experienced by a uniformly aligned and semi-infinite liquid crystal slab. These approximations are treated with more detail in Sec. \ref{sec:boundary}.

\section{Nematic liquid crystals}
\begin{figure}
	\includegraphics[width=0.5\textwidth]{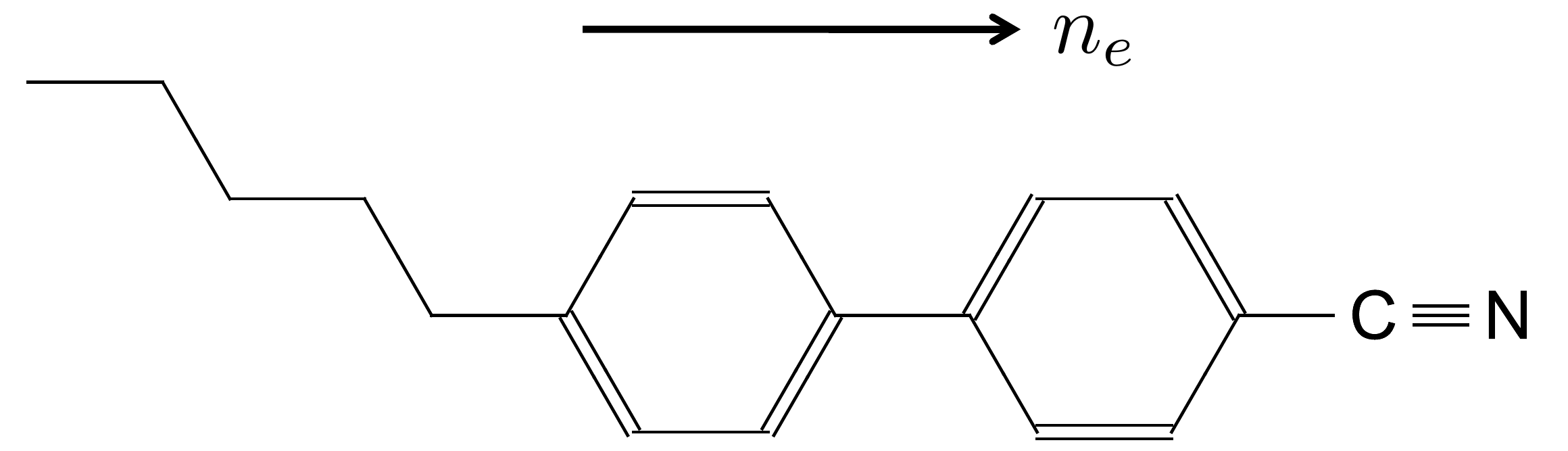}	\caption{\label{fig:molecule}Drawing of the 5CB molecule, which is a nematic liquid crystal at room temperature. The extraordinary axis is along the molecular axis. Because 5CB has positive birefringence, $n_e>n_o$.}
\end{figure}

Most materials undergo a phase transition from solid to liquid at a melting point. Liquid crystals exhibit an intermediate phase in which there is some local ordering due to the molecular structure. Nematic liquid crystals are the simplest class. They can be thought of as long molecules (Fig. \ref{fig:molecule}) that tend to align the long axis with their neighbors. The local direction of molecular orientation is written as a vector of unit length called the director $\bm{n}$. Because the molecules have different dispersive properties along the different axes, one can change the properties of the liquid crystal bulk by manipulating the director.
We carry out calculations using the physical properties of 4-cyano-4$'$-pentylbiphenyl (5CB). In bulk, this is a positive uniaxial material so the dielectric function is higher along the long molecular axis.

\section{Casimir interaction between two infinite slabs}
\begin{figure}
	\includegraphics[width=0.5\textwidth]{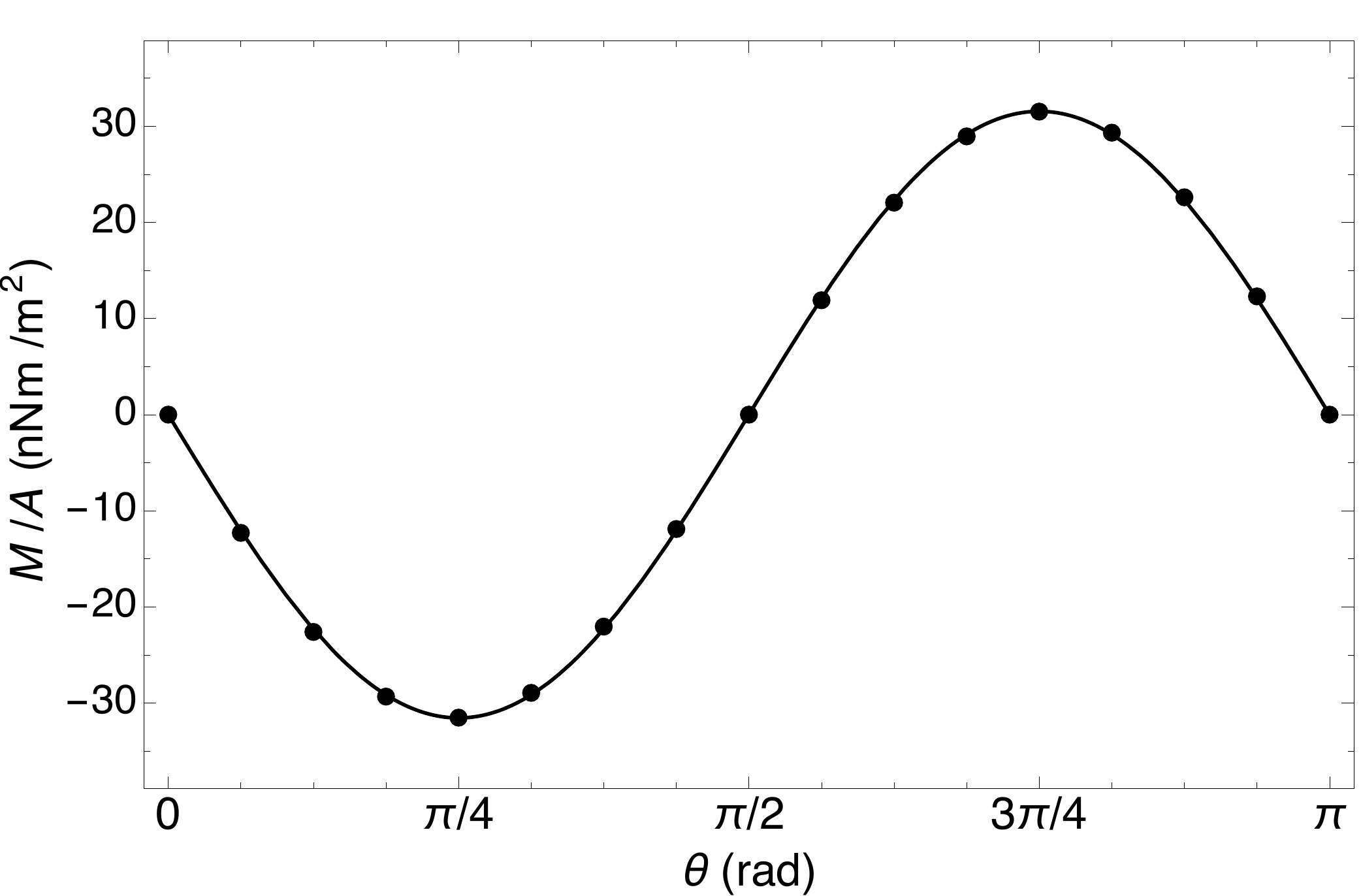}
	\caption{\label{fig:angle_dependence}Calculated torque between BaTiO$_3$ and uniform 5CB bulk separated by 10 nm of SiO$_2$ as a function of angle $\theta$ between the extraordinary axes of the birefringent materials. The line is a fit to a $\sin(2\theta)$ dependence. The difference between the fit and the full calculation is at most 2\%, and is less than 0.1\% for $\theta=\pi/4$.}
\end{figure}

To derive the interaction between two infinite slabs, Barash wrote the Helmhotz free energy between two birefringent materials at finite temperature as an infinite sum over Matsubara frequencies $\xi_n=(2 \pi k_B T /\hbar)n$ \cite{BarashRussian}:
\begin{equation}\label{eq:energy}
\Omega(d,\theta)=\frac{k_B T}{4 \pi^2}\sum_{n=0}^{\infty}{}^{'}\int_0^\infty r dr \int_0^{2\pi}d\varphi\ln{D_n(d,\theta,r,\varphi)},
\end{equation}
where $d$ represents the distance between the two slabs and $\theta$ is the angular separation of their extraordinary axes. We have performed calculations for room temperature based on experimental considerations; however, the zero-temperature calculation for the Casimir torque will provide the correct result to within $<$$10\%$ for the separations presented in this work ($d<50$ nm). The zeros of $D_n(\varphi,r)$ as a function of real frequency $\omega$ indicate allowed surface modes between the two infinite slabs. It can also be expressed in terms of the reflection matrices of the two materials, as in Refs. \cite{TorresGuzman2006,Morgado2013}. In our regime of interest, the energy has $\sin^2{\theta}$ dependence to an excellent approximation, as in Fig. \ref{fig:angle_dependence}. The Casimir torque is then
\begin{equation}\label{eq:torque}
M(d,\theta)=-\frac{\partial \Omega}{\partial \theta}\approx a(d) \sin{2 \theta},
\end{equation}
where $a(d)$ is a negative for the materials considered here (which have positive birefringence). The dielectric functions of the two materials and of the intervening medium are evaluated at imaginary frequencies in Eq. (\ref{eq:energy}). We use the Ninham-Parsegian oscillator model to describe the dispersion of the solid crystals \cite{Mahanty1976}:
\begin{equation}\label{eq:dielectric}
\epsilon(i \xi)=1+\sum_{j=1}^{N}\frac{C_j}{1+\frac{\xi^2}{\omega_j^2}}.
\end{equation}
For the 5CB liquid crystal, we use the dispersive properties at $298.2$ K calculated by Kornilovitch \cite{Kornilovitch2013} using data from Wu \textit{et al.} \cite{Wu1993}. There, the index of refraction is fit with a three-oscillator model, so the dielectric function is:
\begin{equation}\label{eq:5CB}
\epsilon_{\mathrm{5CB}}(i \xi)=1+2\sum_{j=1}^{3}\frac{C_j}{1+\frac{\xi^2}{\omega_j^2}}+ \left ( \sum_{j=1}^{3}\frac{C_j}{1+\frac{\xi^2}{\omega_j^2}} \right ) ^2.
\end{equation}
For the birefringent materials, there are separate functions describing the ordinary and extraordinary axes. The model data used for our calculations are summarized in Table \ref{oscillator}.

\begin{figure}
	\includegraphics[width=0.5\textwidth]{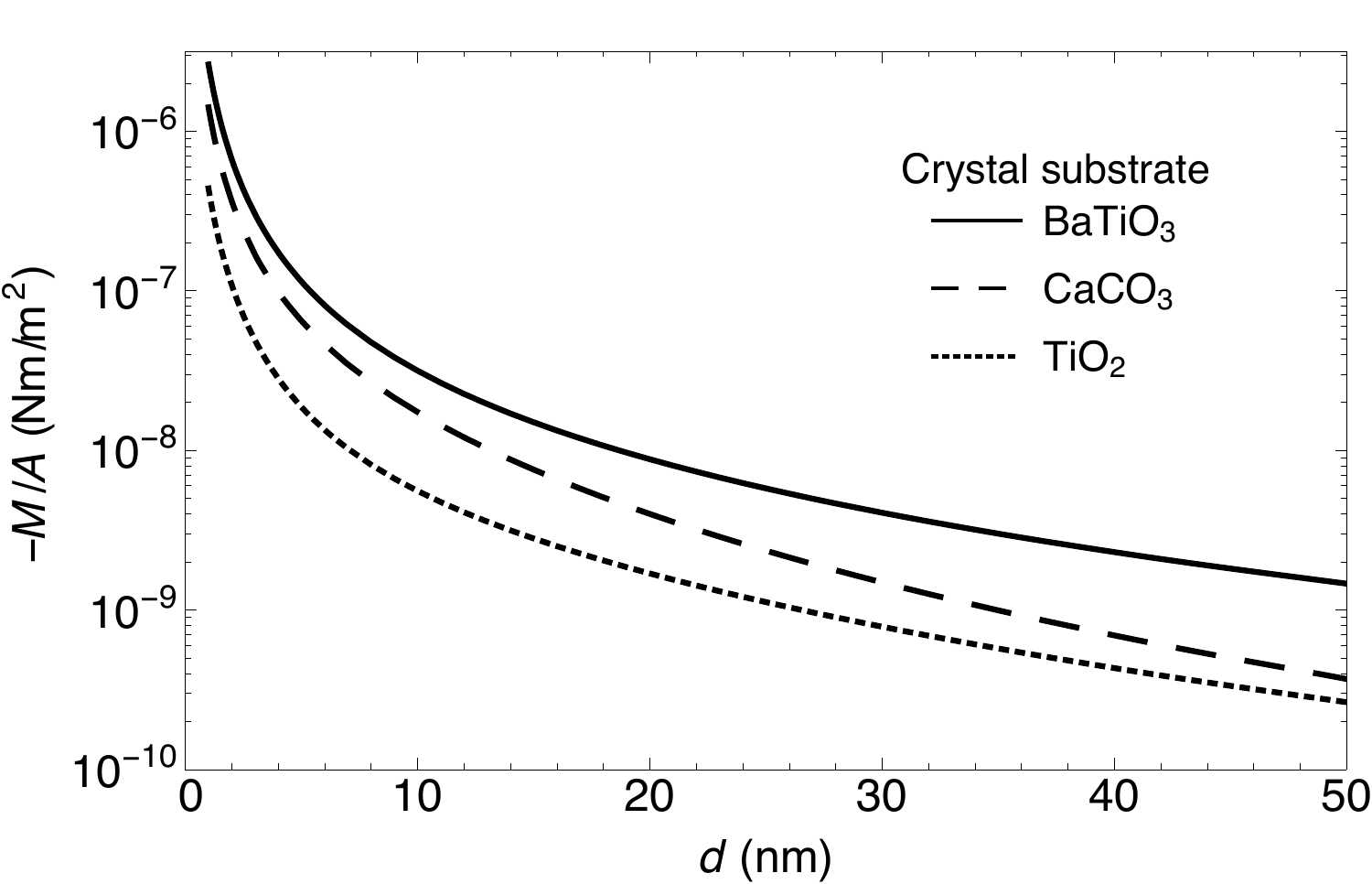}
	\caption{\label{fig:maxtorques}Casimir torque per unit area between a half-slab of aligned 5CB and various birefringent crystals, when separated by a SiO$_2$ layer of thickness \emph{d} with a relative angle of $45^\circ$ between the extraordinary axes.}
\end{figure}

\begin{table}
\caption{Model parameters for dielectric functions of relevant materials. Oscillator data for BaTiO$_3$, CaCO$_3$, TiO$_2$, and SiO$_2$ are from Ref.\cite{Bergstrom1997a}.}
\label{oscillator}
\begin{ruledtabular}
\begin{tabular}{cccccccc}
					&			&	$C_{1}$	&	$\omega_{1}$ (eV)&	$C_{2}$	&$\omega_{2}$ (eV)	&$C_{3,4}$	&	$\omega_{3,4}$ (eV)	\\ \hline
5CB					&	$\perp$	&	0.0374	&	4.40		&	0.1075	&	5.91		&	0.414	&	9.19\\
					&	$||$	&	0.0612	&	4.40		&	0.1025	&	5.91		&	0.460	&	9.19\\
$\mathrm{BaTiO_3}$	&	$\perp$	&	3595	&	0.056		&	4.128	&	5.54		&	--		&	--\\
					&	$||$	&	145.0	&	0.138		&	4.064	&	5.90		&	--		&	--\\
$\mathrm{CaCO_3}$	&	$\perp$	&	1.920	&	0.138		&	1.350	&	13.4		&	--		&	--\\
					&	$||$	&	1.960	&	0.138		&	1.377	&	13.3		&	--		&	--\\
$\mathrm{TiO_2}$	&	$\perp$	&	4.81	&	5.069		&	--		&	--			&	--		&	--\\
					&	$||$	&	5.62	&	4.516		&	--		&	--			&	--		&	--\\
$\mathrm{SiO_2}$	&	--		&	0.829	&	0.057		&	0.095	&	0.099		&	0.798	&	0.133\\
					&			&			&				&			&				&	1.098	&	13.39\\
\end{tabular}
\end{ruledtabular}
\end{table}
We calculate the Casimir torque between an infinite half-space of 5CB liquid crystal and an infinite half-space of several birefringent crystals using Eqs. (\ref{eq:energy}) and (\ref{eq:torque}) and the parameter data in Table \ref{oscillator}. The results for $\theta=45^\circ$ are shown in Fig. \ref{fig:maxtorques}. Note that torques per unit area on the order of 10 nN/m are found for separations of $d=10$ nm.

\section{Torque balance method for measuring liquid crystal anchoring}\label{sec:balance}
The Casimir torque causing a director twist at one boundary competes with the restoring torque from the twisted liquid crystal. The latter is modeled using the Frank free energy density \cite{Frank1958a}. In our geometry, the director is always aligned in the $x$-$y$ plane so its orientation can be written in Cartesian coordinates as $\bm{n}=\{ \cos{\theta(z)}, \sin{\theta(z)},0 \}$. There is no bend or splay of the liquid crystal, so only the twist term contributes to the distortion energy. The twist contribution is given by:
\begin{equation}
F_d=\frac{k_{22}}{2}(\bm{n}\cdot\nabla\times \bm{n})^2,
\end{equation}
where $k_{22}=3.6$ pN is the twist elastic constant of the 5CB\cite{Toyooka1987}. Substituting our expression for $\bm{n}$, the Frank free energy density is
\begin{equation}
F_d=\frac{k_{22}}{2}\left (\frac{\partial \theta}{\partial z}\right )^2,
\end{equation}
and the Frank free energy per unit area is:
\begin{equation}
E_{\mathrm{elastic}}=\int_0^t F_d \mathrm{d}z=\frac{k_{22}}{2}\int_0^t \left (\frac{\partial \theta}{\partial z}\right )^2 \mathrm{d}z,
\end{equation}
where $t$ is the thickness of the liquid crystal layer. In our geometry (as in Fig. \ref{fig:setup}), the extraordinary axis of the solid crystal is along the $x$-axis, so the top boundary at $z=d+t$ (where the liquid crystal is in contact with glass) is treated to induce alignment along $\theta(t+d)=\pi/4$.

A torque applied at $z=d$ twists the director to $\theta(d)=\theta_f$. Using calculus of variations, the lowest energy configuration is given by a linear twist, $\theta(z)=\frac{\pi}{4}+\frac{\Delta \theta}{t}(z-d-t)$, where $\Delta \theta=\pi/4-\theta_f$. The elastic energy of the bulk per unit area is then:
\begin{equation}
E_{\mathrm{elastic}}=\frac{k_{22}}{2}\frac{\Delta \theta^2}{t}.
\end{equation}
If the director at $z=d$ is twisted to $\theta_f$, there is an energy penalty and associated restoring torque at that boundary. The restoring torque of the elastic bulk is given by:
\begin{equation}
M_{\mathrm{elastic}}=-\frac{\partial E_{\mathrm{elastic}}}{\partial \Delta \theta}=-\frac{k_{22}\Delta \theta}{t}.
\end{equation}
$M_\textrm{elastic}$ is the torque that must be applied at the boundary $z=d$ to twist the director to $\theta(d)=\theta_f$. The torque applied at the boundary twists the director until the torque balance equation is satisfied: $M_{\mathrm{elastic}}+M_{\mathrm{external}}=0$. If the Casimir torque is in the approximate form $M_{\mathrm{elastic}}\approx a(d) \sin(2\theta)$, where $\theta$ is the angle between the two extraordinary axes of the birefringent materials, then the torque is approximately $M_{\mathrm{Casimir}}=a(d) \sin(2\theta_f)=a(d) \cos(2\Delta \theta)$ (which has the same form as the planar Rapini-Papoular approximation \cite{Rapini1969}), and the torque balance equation yields:
\begin{equation}\label{eq:balance}
-\frac{k_{22}\Delta \theta}{t}+a(d) \cos{(2\Delta \theta)}=0.
\end{equation}
To predict the director twist for our proposed experiment, we calculate $a(d)$ using Eqs. (\ref{eq:energy}) and (\ref{eq:torque}) and then numerically solve Eq. (\ref{eq:balance}) to find the twist caused by the Casimir torque, $\Delta \theta$. In the proposed experiment, $\Delta \theta$ (or $\theta_f$) will be measured to obtain $a(d)$ via
\begin{equation}
a(d)=\frac{k_{22}}{t} \Delta \theta \sec(2\Delta \theta).
\end{equation}

\section{The boundary layer approximation}
\label{sec:boundary}
The liquid crystal can be treated as an anisotropic bulk material for the calculation because the twisting is small throughout the thickness that experiences the Casimir interaction. The spacing between the liquid crystal and solid birefringent crystal is on the order of $\sim$10 nm, and the liquid crystal layer is about 100  $\mu$m thick. Following Parsegian \cite{Parsegian2006}, the penetration depth of the Casimir interaction is on the order of the material separation. So, the most important region of the liquid crystal is the 10 nm in contact with the SiO$_2$ (or conservatively, the 100 nm). This 100 nm is the 0.1\% of the liquid crystal nearest the birefringent crystal. Because the liquid crystal will be twisted a maximum of $\pi/4$ radians throughout the bulk, the liquid crystal director will vary by a maximum of $\sim $0.05$^\circ$ in the relevant region for the Casimir torque, which has no appreciable effect on the torque's magnitude.

This approximation can also be justified by considering the reflection matrix of the liquid crystal stack. The Casimir energy of the system is a function of the reflection matrices of the two materials at the Matsubara frequencies. Our method assumes that, at the Matsubara frequencies, the reflection matrix of the slowly twisted liquid crystal is nearly the same as that of an untwisted, bulk liquid crystal with the same alignment at the boundary. The first Matsubara frequency at room temperature, $\xi_1=2\pi k_B T/\hbar \approx 245$ THz, corresponds to a wavelength of $\lambda_1 \approx 1$ \textmu m. The higher frequences correspond to shorter wavelengths, so this first term has the longest penetration depth. We calculated the reflection matrices at this frequency between the twisted and untwisted stacks using the Berreman $4\times 4$ matrix method, and found them to be numerically identical to four significant figures \cite{Berreman1972}. The Casimir interaction energy is largely unaffected by the slow twist of the liquid crystal throughout the bulk. Hence, to several significant figures, the torque experienced by the liquid crystal layer is felt entirely at the nearest boundary and is only a function of the director orientation at that boundary.
\section{Proposed experiment}
Common methods for fabricating single liquid crystal cells have been previously reported in Refs. \cite{Miller2013, Waclawik2004} and can be used for this experiment. A rubbed alignment layer of polyvinyl alcohol (PVA)  can be used to cause the liquid crystal molecules to align along the rubbed direction at the surface. The birefringent crystal with a thin, isotropic SiO$_2$ layer (with thickness $d\sim 20$ nm) is sandwiched with the PVA-treated glass with a spacing of $t\approx 100$ $\mu$ m (this value can be measured optically). The liquid crystal is then introduced into the cell via capillary action. The filling process may induce some alignment along the direction of liquid crystal flow; however, baking the sample above the LC clearing temperature ($35^\circ$ C for 5CB) and allowing it to cool slowly will eliminate this effect. As the liquid crystal cools to room temperature, the director settles into the lowest energy state described in Sec. \ref{sec:balance}. The magnitude of the Casimir torque effect can then be measured by observing the twist of the liquid crystal director.

The final director twist $\theta_f$ can be measured optically. This method is similar to a technique for measuring azimuthal surface anchoring strengths of liquid crystals \cite{Faetti1992,Lee2005a}. When linearly polarized light is incident on an adiabatically twisted nematic liquid crystal stack (in which the pitch of the twist is much larger than the wavelength of light), the polarization state is rotated to follow the liquid crystal director. This is known as the adiabatic approximation for twisted nematics and is the principle behind twisted nematic liquid crystal displays \cite{Berreman1972}. In this experiment, white light polarized at $45^\circ$ shines onto the stack as in Fig. \ref{fig:setup}. The Jones vector of this light is $\sqrt{I/2}
\begin{pmatrix}
1\\
1 
\end{pmatrix},$
where $I$ is the intensity. The light polarization follows the director of the twisted nematic and is incident on the transparent SiO$_2$ layer with polarization $\theta_f$. Its Jones vector is now $\sqrt{I}
\begin{pmatrix}
\cos{\theta_f}\\
\sin{\theta_f} 
\end{pmatrix}$.
In a typical measurement of an anchoring force, the liquid crystal is sandwiched between two glass slides that do not interfere with the polarization state of the light, as in Ref. \cite{Faetti2003}. Then, the polarization state $\theta_f$ can be measured with a second polarizer.

\begin{figure}
	\includegraphics[width=0.5\textwidth]{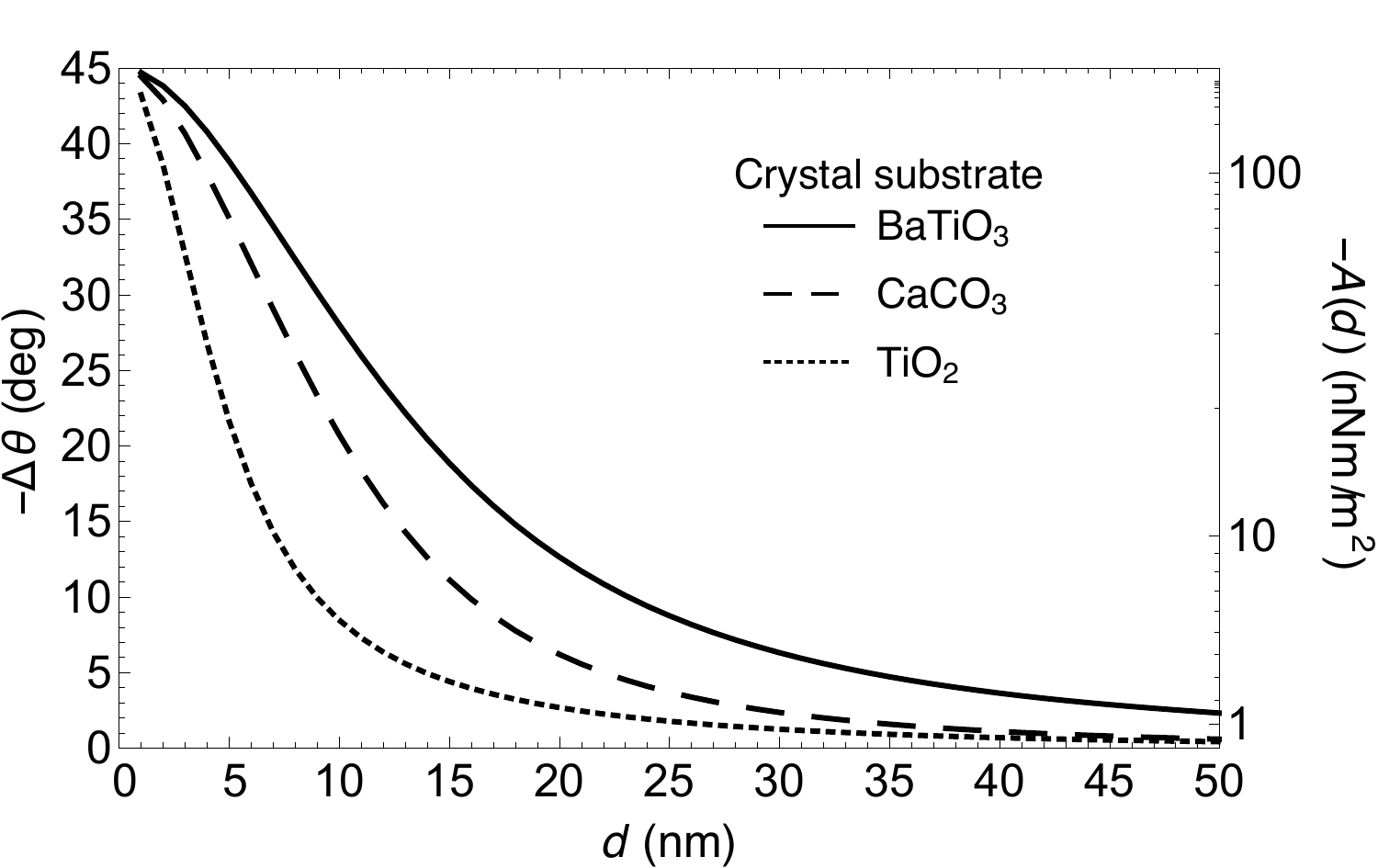}
	\caption{\label{fig:angle_results}Caculated twist of a 100 \textmu m layer of 5CB caused by a Casimir torque induced by various birefringent crystals at distance $d$ from one end of the liquid crystal stack. The incident light is polarized at $45^\circ$ to the ordinary axis at the top of the 5CB stack, but the director is twisted by the Casimir torque, which causes the light polarization to rotate $\Delta \theta$ towards the extraordinary axis at $0^\circ$.}
\end{figure}
To calculate the expected results from an experiment, we consider a liquid crystal layer thickness of 100 $\mu$m and calculate the lowest energy state of the system using Eq. (\ref{eq:balance}). Figure \ref{fig:angle_results} shows the expected results for this case. The liquid crystal bulk is predicted to twist by over $35^\circ$ when the stack is separated by 5 nm from BaTiO$_3$, and a twist of several degrees is expected for separations of $d \sim 50$ nm (well into the Casimir regime). When near a birefringent material that has negative birefringence over a large frequency range (such as lithium niobate), the liquid crystal would twist towards the ordinary axis instead of the extraordinary axis. This would provide further confirmation that dispersion effects are causing the director to twist.

\section{conclusions}
We have proposed an experiment for measuring a Casimir torque between a birefringent crystal and a liquid crystal separated by an isotropic spacer layer. We provide complete calculations of the expected results for several materials at a range of separations and include details for a proposed experiment. This experimental design avoids many of the difficulties involved with a torsion pendulum or levitating microdisks.

\end{document}